\documentstyle[12pt,aaspp4,epsf]{article}
\newcommand{\beq}{\begin{equation}}
\newcommand{\beqa}{\begin{eqnarray}}
\newcommand{\eeq}{\end{equation}}
\newcommand{\eeqa}{\end{eqnarray}}
\newcommand{\simg}{\gtrsim}
\newcommand{\siml}{\lesssim}

\begin{document}
%
\title{ 
Microlensing of collimated Gamma-Ray Burst afterglows
}
\author{
Kunihito Ioka$^{1}$
and Takashi Nakamura$^{2}$
}
\affil{$^{1}$ Department of Physics, Kyoto University, Kyoto 606-8502,
Japan}
\affil{
$^{3}$Yukawa Institute for Theoretical Physics, Kyoto University, 
Kyoto 606-8502, Japan}
\centerline{Dec~~2~~2000}
\authoremail{iokakuni@tap.scphys.kyoto-u.ac.jp}

\begin{abstract}
We investigate stellar microlensing of 
the collimated gamma-ray burst afterglows.
A spherical afterglow appears on the sky as a superluminally 
expanding thin ring (``ring-like'' image), 
which is maximally amplified as it crosses the lens.
We find that the image of the collimated afterglow 
becomes quite uniform (``disk-like'' image)
after the jet break time
(after the Lorentz factor of the jet drops below the inverse of 
the jet opening angle).
Consequently, the amplification peak in the light curve 
after the break time is lower and broader.
Therefore detailed monitoring of the amplification history
will be able to test whether the afterglows are jets or not,
i.e., ``disk-like'' or not,
if the lensing occurs after the break time.
We also show that some proper motion and polarization is expected,
peaking around the maximum amplification.
The simultaneous detection of the proper motion and the polarization
will strengthen that the brightening of the light curve
is due to microlensing.
\end{abstract}

\keywords{gamma rays: bursts --- gravitational lensing}

\section{INTRODUCTION}
Whether the gamma-ray bursts (GRBs) 
are collimated or not is one of the most important questions in GRBs,
since the jet configuration is crucial for almost all aspects of GRBs,
such as the energetics, statistics, and central engine models of GRBs.
There are several suggestions for the collimation of GRBs.
When the Lorentz factor of the ejecta drops below the inverse of the 
opening angle, the ejecta begins to spread sideways,
and we expect a break in the light curve of the afterglow (Rhoads 1999).
Such breaks have been observed in several GRBs,
such as GRB 990510 (Stanek et al. 1999; Harrison et al. 1999)
and GRB 991216 (Halpern et al. 2000).
The rapid decline rate (e.g., GRB 980519) 
or the unusually slow decline rate of GRB 980425 also suggests
the collimation of GRBs 
(Halpern et al. 1999; Sari, Piran \& Halpern 1999; Nakamura 1999).
However, in order to establish that the afterglows are jets,
other observations are indispensable, such as
early afterglow observations at radio frequencies (Frail et al. 2000).
Microlensing of the afterglows has a potential to be one of such observations.
The emission region of the afterglow seen by an external observer
occupies a size of $\sim 10^{17}$ cm on the sky
which is comparable to the lensing zone of a solar mass lens
located at the cosmological distance (Loeb \& Perna 1998).
Therefore microlensing can resolve the afterglow,
and hence we can expect that some features of jets can be obtained 
by the observations of the microlensed afterglows.

On the other hands, afterglows will be useful to 
constrain the cosmological density parameter 
of stellar mass objects $\Omega_{\star}$ 
through microlensing.
As a result of the relativistic motion,
the emission region appears to expand superluminally,
so that the microlensing time is only $\sim 1$ day (Loeb \& Perna 1998).
This is much shorter than that of common sources,
$\simg 10$ years (Gould 1995).
The microlensing probability of a source at a 
redshift $z_s \sim 3$ is $\sim 0.3 \Omega_{\star}$
(Press \& Gunn 1973; Gould 1995; Koopmans \& Wambsganss 2000).
Since the known luminous stars amount to 
$\Omega_{\star} \simg 4 \times 10^{-3}$ (Fukugita, Hogan \& Peebles 1998),
roughly 1 in $\sim 10^3$ GRBs would be within the lensing zone.
If some fraction of the dark matter is made of massive
compact halo objects (MACHOs), as Galactic microlensing searches suggest 
(Alcock et al. 2000), 
we can expect higher probability.
Remarkably, the microlensed afterglow may have been already observed
in the light curve of GRB 000301C as the achromatic bump
(Garnavich, Loeb \& Stanek 2000; Gaudi, Granot \& Loeb 2001;
but see also Panaitescu 2001).
Since new GRB satellites such as {\it HETE-2} and {\it SWIFT}
will provide many opportunities to monitor afterglows frequently,
it is important to study the jet effects
on the microlensing of the GRB afterglows.

In \S 2 and 3, we will calculate the equal arrival time surfaces,
and the surface brightness distribution of the afterglow image,
respectively, taking into account of the jet effects.
In \S 4, we will calculate the microlensed light curve.
In \S 5, we will investigate the proper motion and the polarization
induced by microlensing.
\S 6 is devoted to summary and discussions.

\section{EQUAL ARRIVAL TIME SURFACE}
Let $E$, $\rho$, $\theta_i$ and $c_s$ be the burst energy, 
the density of the ambient gas, 
the initial opening angle,
and the sound speed of the jet, respectively.
The Lorentz factor of the shock front approximately evolves as 
\beqa
\Gamma=\left\{
\begin{array}{ll}
\Gamma_b \left({{r}/{r_b}}\right)^{-3/2},& \quad r<r_b,\\
\Gamma_b \exp \left\{-(r-r_b)/r_{\Gamma}\right\},& \quad r>r_b,
\end{array}\right.
\label{eq:gamma}
\eeqa
where $r_b=(75 E/8 \pi \rho c_s^2)^{1/3}=3.9 \times 10^{18} 
E_{52}^{1/3} n_1^{-1/3} (\sqrt{3} c_s/c)^{-2/3} {\rm cm}$,
$r_{\Gamma}=(8/75)^{1/3} r_b$ and $\Gamma_b=2 c_s/5 c \theta_i$
(Rhoads 1999).
Here $E_{52}$ is the burst energy in units of $10^{52}$ ergs,
and $n_1$ is the ambient gas density in ${\rm cm}^{-3}$.
We assume that the material is uniformly distributed
across the jet at any $r$,
the emission comes just behind the shock front, and $\Gamma \gg 1$.
Such approximations are sufficient and convenient to understand 
the key features before performing more detail calculations
(Rhoads 1999; Panaitescu \& M${\acute {\rm e}}$sz${\acute {\rm a}}$ros 1999;
Moderski, Sikora \& Bulik 2000; Huang, Dai \& Lu 2000),
although the future detailed analysis may demand the realistic model
(Granot, Piran \& Sari 1999; Granot \& Loeb 2001; Gaudi, 
Granot \& Loeb 2001).

The arrival time $T$ of a photon emitted at an angle $\theta$
from the line of sight is given by
$T=t-r \mu/c$ where $t=\int dr /c \sqrt{1-\Gamma^{-2}}$
and $\mu=\cos \theta$.
This relation with equation (\ref{eq:gamma}) can be written as
\beqa
1-\mu=\left\{
\begin{array}{ll}
{{1}\over{8\Gamma_b^2}}\left({{r}\over{r_b}}\right)^{-1}
\left[{{T}\over{T_b}}-\left({{r}\over{r_b}}\right)^4\right]
,& \quad r<r_b,\\
{{1}\over{8\Gamma_b^2}}\left({{r}\over{r_b}}\right)^{-1}
\left[{{T}\over{T_b}}-1-{{2 r_{\Gamma}}\over{r_b}} 
\left[\exp \left\{{{2(r-r_b)}\over{r_{\Gamma}}}\right\}-1\right]\right]
,& \quad r>r_b,
\end{array}\right.
\label{eq:equals}
\eeqa
where $T_b=r_b/8 c \Gamma_b^2=0.35 E_{52}^{1/3} n_1^{-1/3}
(\sqrt{3} c_s/c)^{-8/3} (\theta_i/0.01)^{2} {\rm days}$.
Equation (\ref{eq:equals}) gives the equal arrival time surface
at $T$, and some examples are shown in Figure 1.
In the power law regime, $T \siml T_b$,
the apparent size $r_{\perp,max}$ of the jet is 
determined by the equal arrival time surface
(Sari 1998; Panaitescu \& M${\acute {\rm e}}$sz${\acute {\rm a}}$ros 1998).
The distance of the equal arrival time surface from the line of 
sight is given by
$r_{\perp}=r \sqrt{1-\mu^2} \simeq r \sqrt{2(1-\mu)}$
with equation (\ref{eq:equals}),
which takes the maximum value of
$r_{\perp,max}=(T/5 T_b)^{5/8} r_b/\Gamma_b=
1.2 \times 10^{17} E_{52}^{1/8} n_1^{-1/8}
(\theta_i/0.01)^{-1/4} (T/1 {\rm day})^{5/8} {\rm cm}$ 
at $r/r_b=(T/5 T_b)^{1/4}$.
On the other hand, in the exponential regime, $T \simg T_b$, 
the apparent size of the jet is determined by
the opening angle of the jet,
$\theta_b=\theta_i+c_s t_{co}/r$, where $t_{co}=\int dr/c \Gamma$
is the comoving time in the jet frame.
With equation (\ref{eq:gamma}), we find
\beqa
{{\theta_b}\over{1/\Gamma_b}}=\left\{
\begin{array}{ll}
{{2}\over{5}}{{c_s}\over{c}}
\left[1+
\left({{r}\over{r_b}}\right)^{3/2}\right]
,& \quad r<r_b,\\
{{2}\over{5}}{{c_s}\over{c}}
\left[1+\left({{r}\over{r_b}}\right)^{-1}
\left\{1+{{5 r_{\Gamma}}\over{2 r_b}}
\left[\exp\left\{{{(r-r_b)}\over{r_{\Gamma}}}\right\}-1\right]
\right\}\right]
,& \quad r>r_b.
\end{array}\right.
\label{eq:thetab}
\eeqa
If the direction of the jet axis is $(\theta, \phi)=(\theta_0,0)$
with $\theta=0$ being the line of sight,
the edge of the jet is given by 
$\theta^2+\theta_0^2-2 \theta \theta_0 \cos \phi=\theta_b^2$.
In Figure 1, the trajectories of jet edges for $\theta_0=0$ and 
$\theta_0=\theta_i$ on the $\phi=0$ plane are plotted.
We can see that there is an intermediate regime
where the evolution of the Lorentz factor is power law
while the edge of the jet determines the apparent size
(Rhoads 1999; Panaitescu \& M${\acute {\rm e}}$sz${\acute {\rm a}}$ros 1999).

\section{SURFACE BRIGHTNESS DISTRIBUTION}
Let us consider a system that has an isotropic luminosity $d L_\nu$
in its rest frame and is moving with Lorentz factor $\gamma$
in a direction $(\theta, \phi)$ with a solid angle $d\Omega=d\mu d\phi$.
Emitted photon with frequency $\nu$ is blueshifted to
frequency $\nu/\gamma(1-\beta \mu)$, where $\beta=\sqrt{1-\gamma^{-2}}$.
Using the Lorentz transformation,
the observed flux at distance $D$ and frequency $\nu$ is given by
$d F_{\nu}={{d L_{\nu \gamma (1-\beta \mu)}}/
{4 \pi D^2 \gamma^3 (1-\beta \mu)^3}}$.
A jet, with a luminosity $L_{\nu}$ in its local frame,
is a collection of such systems, so that
we have $d L_{\nu}=L_{\nu} d\Omega/\pi \theta_b^2$
with equation (\ref{eq:thetab}).
Now we assume that the luminosity in the local frame is
$L_{\nu} \propto \Gamma^a \nu^b$.
Then, 
the flux per unit area, i.e., the surface brightness
$S(r_{\perp}, \phi; T):={{d^2 F_{\nu}}/{r_\perp d r_\perp d\phi}}$
can be obtained by
\beq
S(r_{\perp}, \phi; T) \propto \Gamma^{a+b-3} (1-\beta \mu)^{b-3}
\theta_b^{-2} {{d\mu}\over{dr}} \left/ 
\left|{{d r_{\perp}^2}\over{dr}}\right|\right.,
\label{eq:sbd}
\eeq
where we use $d\Omega/r_\perp d r_\perp d\phi=2 d\mu/d r_\perp^2$
and the fact that the Lorentz factor of the shock front $\Gamma$
is higher than that of the material behind it $\gamma$
by a factor of $\sqrt{2}$ (Blandford \& Mckee 1976).
With equations (\ref{eq:gamma}), (\ref{eq:equals})
and (\ref{eq:thetab}), we can find that equation (\ref{eq:sbd})
is a function of $r$, noting $r_{\perp} \simeq r \sqrt{2(1-\mu)}$ and
$1-\beta\mu \simeq 1-\beta +1-\mu
\simeq 1/\Gamma^2+1-\mu$.
Since we can obtain $r$ for given $r_{\perp}$ by 
$r_{\perp} \simeq r \sqrt{2(1-\mu)}$ with equation (\ref{eq:equals}), 
the surface brightness can be calculated as a function of $r_\perp$.
Note that there are two solutions of $r$ for each $r_\perp$,
one from the front of the equal arrival time surface
and the other from its back.
As time goes, the emission from the back does not contribute
due to the jet geometry, as we can see from Figure 1.

The assumption $L_{\nu} \propto \Gamma^a \nu^b$ is valid
for frequencies far from the typical synchrotron frequency $\nu_m$
(Sari, Piran \& Narayan 1998; Sari, Piran \& Halpern 1999).
For simplicity, we neglect scattering, self-absorption, and electron cooling.
Then, the luminosity is proportional to the total number of 
swept-up electrons $N_e \propto \Gamma^{-2}$ (Rhoads 1999) 
times the radiation power from each electron 
$P\propto \gamma^2 B^2 \propto \Gamma^4$.
Since the typical frequency is $\nu_m\propto B \gamma^2 \propto \Gamma^3$,
the luminosity at the typical frequency is $L_{\nu_m}\propto \Gamma^{-1}$.
At $\nu \ll \nu_m$, 
we have $L_\nu=L_{\nu_m}(\nu/\nu_m)^{1/3}\propto \Gamma^{-2} \nu^{1/3}$.
At $\nu \gg \nu_m$, if electrons are accelerated
to a power law distribution with index $p$,
we have
$L_\nu=L_{\nu_m}(\nu/\nu_m)^{-(p-1)/2}\propto 
\Gamma^{(3p-5)/2} \nu^{-(p-1)/2}$.
Therefore, we find
$a=-2$, $b=1/3$ at $\nu \ll \nu_m$, and
$a=(3p-5)/2$, $b=-(p-1)/2$ at $\nu \gg \nu_m$.
We shall adopt $p=2.5$.

In Figure 2, we show the surface brightness distribution as
a function of the distance from the center
for $\theta_0=0$ (the jet axis coincides with the line of sight).
In the power law regime $T/T_b \siml 1$, the surface brightness
is brighter near the edge and dimmer at the center, i.e.,
``ring-like'' (Waxman 1997; Sari 1998;
Panaitescu \& M${\acute {\rm e}}$sz${\acute {\rm a}}$ros 1998).
However, in the exponential regime $T/T_b \simg 1$,
the surface brightness becomes nearly constant, i.e.,
``disk-like''.
The qualitative feature does not depend on $\theta_0$.
In the intermediate regime, a part of the ring will be missing
due to the jet edge when $\theta_0 \ne 0$.

\section{MICROLENSED LIGHT CURVE}
We now consider a point mass lens of mass $M$ that
is located at $L$ from the source center on the sky
(see Mao \& Loeb 2001 for binary lenses).
The Einstein radius on the source plane
is given by $R_E=[(4 G M/c^2)(D_s D_{ls}/D_l)]^{1/2}$,
with $D_l$, $D_s$ and $D_{ls}$ being the angular diameter
distance to the lens, to the source, and from the lens to the source,
respectively (e.g., Schneider, Ehlers \& Falco 1992).
A point lens is described by three parameters,
the Einstein radius,
$r_E:={{R_E}/{(r_b/\Gamma_b)}}=0.45
\left({{M}/{M_{\odot}}}\right)^{1/2}
\left[({D_s D_{ls}/D_l})/{10^{28} {\rm cm}}\right]^{1/2}
E_{52}^{-1/3} n_1^{1/3} 
\left({{\sqrt{3} c_s}/{c}}\right)^{5/3}
\left({{\theta_i}/{0.01}}\right)^{-1}$,
the impact parameter, $l:=L/(r_b/\Gamma_b)$,
and the azimuthal angle of the lens position, $\phi_l$,
with respect to the source center on the sky.
Here we measure the distance from the line of sight
in units of $r_b/\Gamma_b$.

The observed flux can be calculated numerically by
$F_{\nu}={\int A(r_{\perp}, \phi) 
S(r_{\perp}, \phi; T) r_{\perp} d r_{\perp} d\phi}$,
where $A(r_{\perp}, \phi)=(u_l^2+2)/u_l \sqrt{u_l^2+4}$
and $u_l=[L^2+r_{\perp}^2-2 L r_{\perp} \cos (\phi-\phi_l)]^{1/2}/R_E$.
The unlensed flux can be obtained by putting $A=1$.
In Figure 3, the lensed flux of an afterglow,
the unlensed flux, and the lensed flux 
while retaining the initial surface brightness
(i.e., retaining the ``ring-like'' image)
are plotted by solid, dashed, and dotted lines, respectively.
Only the exponential regime $T \simg T_b$ is depicted,
where the slope of the unlensed light curve 
is $-p=-2.5$ at $\nu \gg \nu_m$,
and $-1/3$ at $\nu \ll \nu_m$ (Sari, Piran \& Halpern 1999).
At early times, the temporal profiles of the lensed and unlensed flux
have the same shape but different amplitudes,
since the source can be regarded as pointlike.
The offset between the lensed and unlensed flux is a function of $l/r_E$.
The maximum amplification occurs around
when the edge of the image crosses the lens, 
$r_{\perp,max} \sim L=l r_b/\Gamma_b$.
Here we set $l=1$ so that the the maximum amplification occurs at 
$T \sim 10 T_b$, with the image being ``disk-like''.

In the upper panel of Figure 3, 
we consider the case $\nu \gg \nu_m$.
At the maximum amplification,
an amplification peak would appear
if the image were ``ring-like'',
as in the power law regime\footnote{
This statement may depend on the more detail profile 
of the ring (Granot et al. 1999).} (Loeb \& Perna 1998).
However, since the image is ``disk-like'',
the amplification peak becomes lower and broader,
and apparently
disappears for $\nu \gg \nu_m$.
While for $\nu \ll \nu_m$, as shown in the lower panel of Figure 3,
the low amplification peak does exit in the light curve
even if the image is ``disk-like''.
This is because the decline rate of the unlensed flux at $\nu \gg \nu_m$ 
is too steep ($-p=-2.5$) for the time derivative of the flux to be positive,
but the slope at $\nu \ll \nu_m$ is very mild ($-1/3$).
Note that the amplification factor is nearly the same in both cases.
In Figure 4, the amplification factor
(while retaining the initial surface brightness)
is plotted by solid (dotted) lines.
The amplification factor for the uniform ring with a fractional 
width 10\%, which reproduces the bump in the light curve of GRB 000301C
(Garnavich, Loeb \& Stanek 2000), is also plotted by dashed lines.
Errorbars of $\pm 0.1$ magnitude are also shown for reference.
As analyzed by Gaudi \& Loeb (2001) and Gaudi, Granot \& Loeb (2001),
a future monitoring campaign of a lensed afterglow
will be able to reconstruct the radial structure 
of the afterglow image,
although the quality of the observational data for GRB 000301C
(about $\sim \pm 0.1$ magnitude) may not be sufficient for this purpose.
If the reconstructed image is "disk-like" after the break time,
it will be strengthened that the afterglows are jets.

In the intermediate regime,
where the evolution of the Lorentz factor is power law
but the edge of the jet determines the image size,
the amplification depends on the lens position $\phi_l$
when $\theta_0 \ne 0$.
If the lens is located on the side where 
the ring is cut off by the jet edge,
the amplification peak is small, while,
if the lens is on the other side,
the peak is relatively large.

\section{PROPER MOTION AND POLARIZATION}
Even when the multiple images formed by microlensing can not be resolved,
the centroid of the combined image is expected to move
(Hosokawa et al. 1993, H\o g, Novikov \& Polnarev 1995,
Walker 1995, Miyamoto \& Yoshii 1995, Paczy${\acute {\rm n}}$ski 1998,
Mao \& Witt 1998).
The order of the maximum displacement is estimated by 
$R_E/D_s \sim 10^{-11} (M/M_{\odot})^{1/2} 
\sim 2 (M/M_{\odot})^{1/2} \mu{\rm as}$ 
with $R_E \sim 10^{17} (M/M_{\odot})^{1/2} {\rm cm}$
and $D_s \sim 10^{28} {\rm cm}$.
This proper motion might be measured by the upcoming missions,
such as the Space Interferometry Mission (SIM; see http://sim.jpl.nasa.gov),
with positional accuracy down to $\sim 1 \mu{\rm as}$
(Paczy${\acute {\rm n}}$ski 1998).
To neglect the proper motion due to the jet edge effect (Sari 1999),
we set here $\theta_0=0$.
We can calculate the light centroid numerically by weighting the image
positions with brightness (Walker 1995, Mao \& Witt 1998).
The upper panel of Figure 5 shows the proper motion as a function of time
at $\nu \gg \nu_m$.
Initially, since the source is pointlike,
the initial displacement is a function of $l/r_E$
(see Figure 1 in Walker 1995).
As the source expands, the centroid moves toward the lens
due to the finite source effect (Mao \& Witt 1998).
The centroid goes through the origin in coincidence with
the maximum amplification.
The upper panel of Figure 6 shows
the maximum magnitude of the displacement as a function of
the normalized impact parameter $l/r_E$
for the uniform surface brightness (thick lines),
the initial surface brightness (dotted lines)
and the uniform ring with a fractional width 10\% (dashed lines).
Since the surface brightness is retained,
the displacement depends only on $l/r_E$.
The displacement is smaller for more uniform surface brightness
but is always larger than the initial displacement.

We also expect some polarization in the microlensed afterglows
(Loeb \& Perna 1998).
Here, following Sari (1999), we shall consider the favorite
conditions in which
the polarization at each point in the image is 
$\Pi_0=(p+1)/(p+7/3)\simeq 72 \%$ toward the source center.
To neglect the polarization due to the jet edge effect
(Ghisellini \& Lazzati 1999; Sari 1999),
we set here $\theta_0=0$.
The net polarization can be calculated by
averaging $\Pi_0 \cos(2 \phi)$ with $A(r_{\perp}, \phi) S(r_{\perp}, \phi; T)$.
The lower panel of Figure 5 shows the net polarization as a function of time
at $\nu \gg \nu_m$.
The polarization has a maximum toward the lens
in coincidence with the maximum amplification.
The lower panel of Figure 6 shows
the maximum polarization as a function of
the normalized impact parameter $l/r_E$
for the uniform surface brightness (thick lines),
the initial surface brightness (dotted lines)
and the uniform ring with a fractional width 10\% (dashed lines).
Even in the exponential regime,
the polarization due to microlensing is comparable to the polarization 
which is expected around the jet break time
when $\theta_0 \ne 0$ (see Figure 4 in Sari 1999).

\section{SUMMARY AND DISCUSSIONS}
We have investigated stellar microlensing of collimated GRB afterglows
taking into account of the jet effects.
Using the analytical expressions for the evolution of the jet,
we have first calculated the surface brightness distribution of the jet.
We find that the image is ``disk-like'' after the jet break time,
rather than ``ring-like'' before the break time.
We have further analyzed the microlensing signal 
in the light curve of the afterglow observed at frequencies 
far below and far above the typical synchrotron frequency $\nu_m$, where
the radio is below $\nu_m$ for about 1 month,\footnote{
The radio scintillation is suppressed at $\nu \simg 10$ GHz (Goodman 1997).}
and the optical is above $\nu_m$ after about 1 day.
If the edge of the image crosses the lens 
before the break time, the microlensing signal appears as
an achromatic amplification peak in the light curve (Loeb \& Perna 1998).
The peak seen in GRB 000301C is believed to be a microlensing event,
occurred before the break time, in the ``ring-like'' regime
(Garnavich, Loeb \& Stanek 2000; Gaudi, Granot \& Loeb 2001;
but see also Panaitescu 2001).
We find that, after the break time,
the amplification peak becomes lower and broader
because of the ``disk-like'' image.
Since detailed monitoring of the amplification history 
will be able to reconstruct the afterglow image 
(Gaudi \& Loeb 2001; Gaudi, Granot \& Loeb 2001),
it could be tested whether the afterglows are jets or not, i.e.,
the reconstructed image is "disk-like" or not after the break time.
We should take care that
the amplification peak apparently disappears for $\nu \gg \nu_m$.
At the break time, the peak amplitude
depends on the lens position.

Microlensing also induces the proper motion 
and the polarization of the afterglow
in coincidence with the maximum amplification.
The magnitude of the proper motion
is order of the Einstein radius of the lens,
which might be detected by upcoming missions,
such as SIM, with positional accuracy down to $\sim 1 \mu{\rm as}$.
{}From the proper motion 
we can estimate the apparent size of the afterglow explicitly.
The maximum polarization due to microlensing is comparable to 
the polarization which is expected around the jet break time (Sari 1999),
even if the image is ``disk-like''.
The simultaneous detection of the proper motion and the polarization
will strengthen that the brightening of the light curve
is due to microlensing,
although an initial constant positive offset in the amplification history
(Loeb \& Perna 1998) and
a specific level of chromaticity to the amplification history
(Granot \& Loeb 2001)
are unique features of the lensed afterglows.


\acknowledgments
We would like to thank the anonymous referee 
for useful comments to improve our paper.
We would like to thank H. Sato, S. Kobayashi and T. Chiba for
useful discussions. 
This work was supported in part by
Grant-in-Aid for Scientific Research Fellowship
of the Japanese Ministry of Education,
Science, Sports and Culture, No.9627 (KI) and by
Grant-in-Aid of Scientific Research of the Ministry of Education,
Culture, and Sports, No.11640274 (TN) and 09NP0801 (TN).


%
%

\newpage 
\begin{figure}
    \epsfysize 17cm 
    \epsfbox{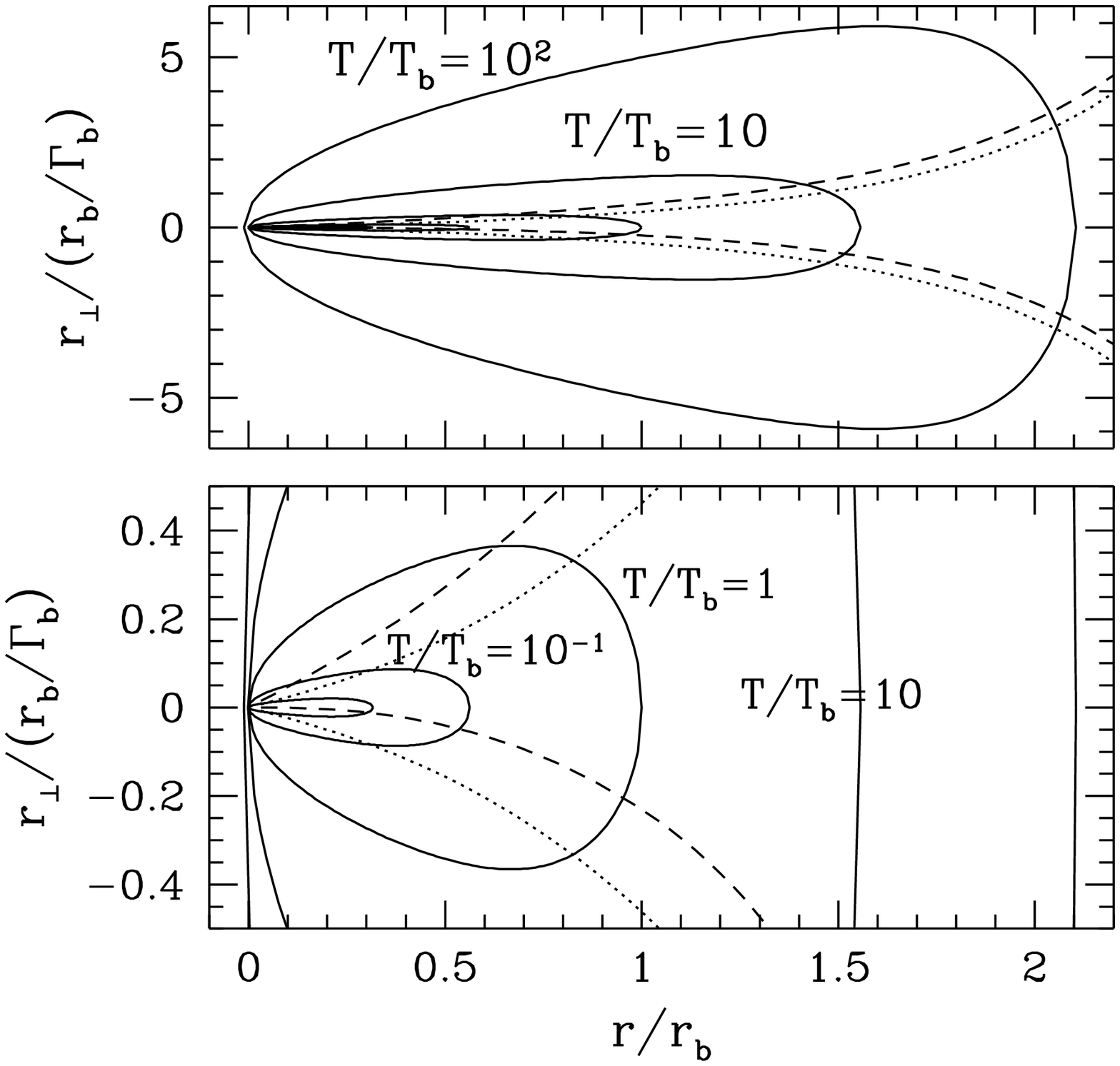}
\caption[fig1.ps]
{Equal arrival time surfaces, i.e., the surfaces from which
photons from a thin shell arrives at the same time $T$
to the observer far on the right side, are shown by solid lines.
We adopt $\theta_i=0.01$ and $c_s=c/\sqrt{3}$.
The upper and lower panels are the same, but the scale of the vertical axis
is different.
The arrival times are $T/T_b=10^{-2}, 10^{-1}, 1, 10, 10^2$
as the surface gets larger.
The Lorentz factor of the emitting shell obeys equation (\ref{eq:gamma}),
in which the shell decelerates exponentially at $r>r_b$.
The trajectories of the jet edges
for $\theta_0=0$ and $\theta_0=\theta_i$ are also plotted
with dotted and dashed lines, respectively,
with $\theta_0=0$ being the line of sight.
No photon comes from the outside of these trajectories.
}
\end{figure}

\newpage 
\begin{figure}
    \epsfysize 17cm 
    \epsfbox{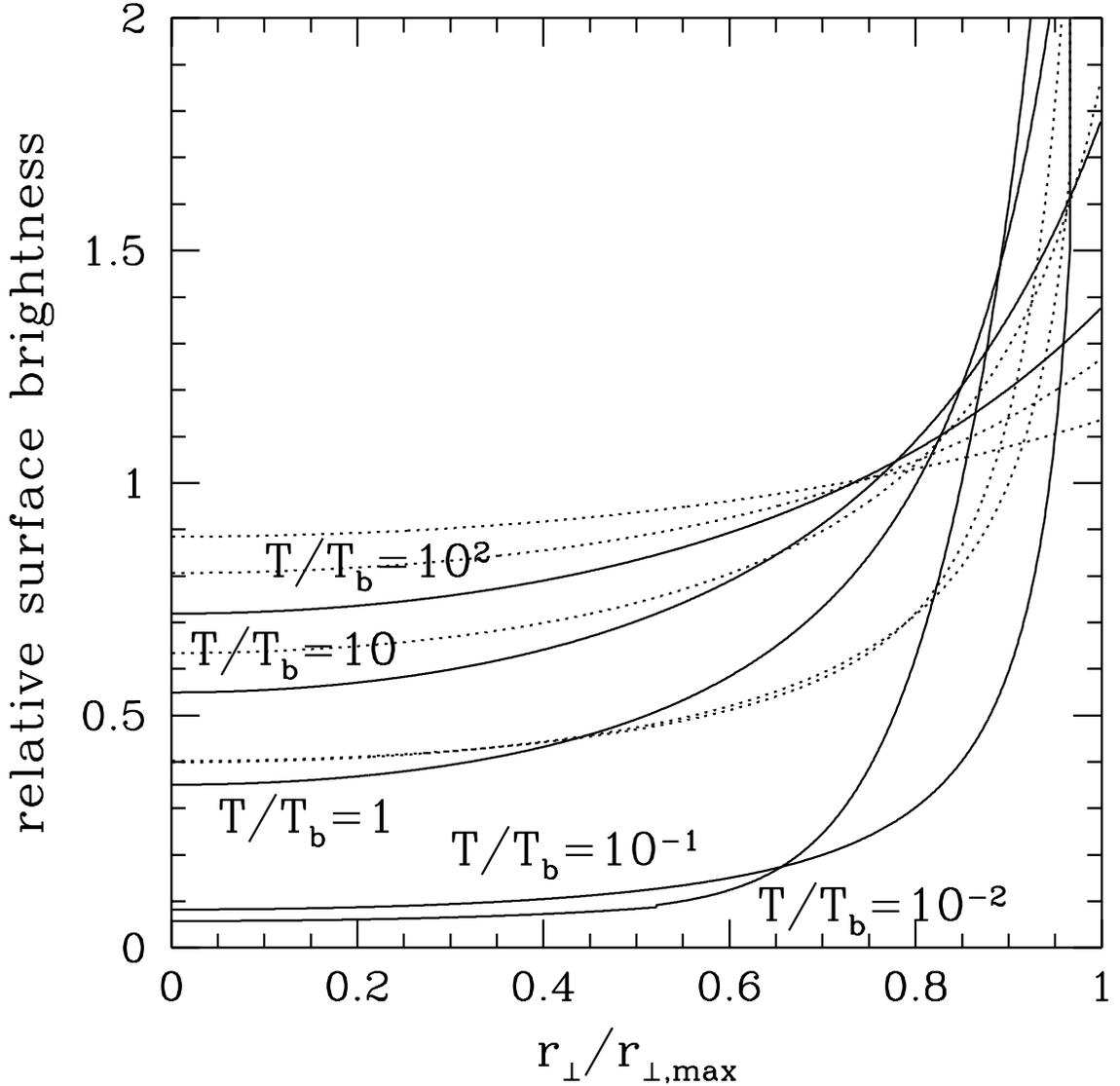}
\caption[fig2.ps]
{The surface brightness as a function of the distance from the center
at the times $T/T_b=10^{-2}, 10^{-1}, 1, 10, 10^2$ is shown
for frequencies above and below the typical
synchrotron frequency with solid and dotted lines,
respectively.
Each surface brightness is normalized by the average surface brightness.
The relative brightness at the center increases with the time $T$.
The image is ``ring-like'' at $T \siml T_b$ and ``disk-like'' at 
$T \simg T_b$.
The divergence of the brightness at $r_{\perp}=r_{\perp, max}$
is an artifact of the assumption that the radiation comes from
a two-dimensional shell.
}
\end{figure}

\newpage 
\begin{figure}
    \epsfysize 17cm 
    \epsfbox{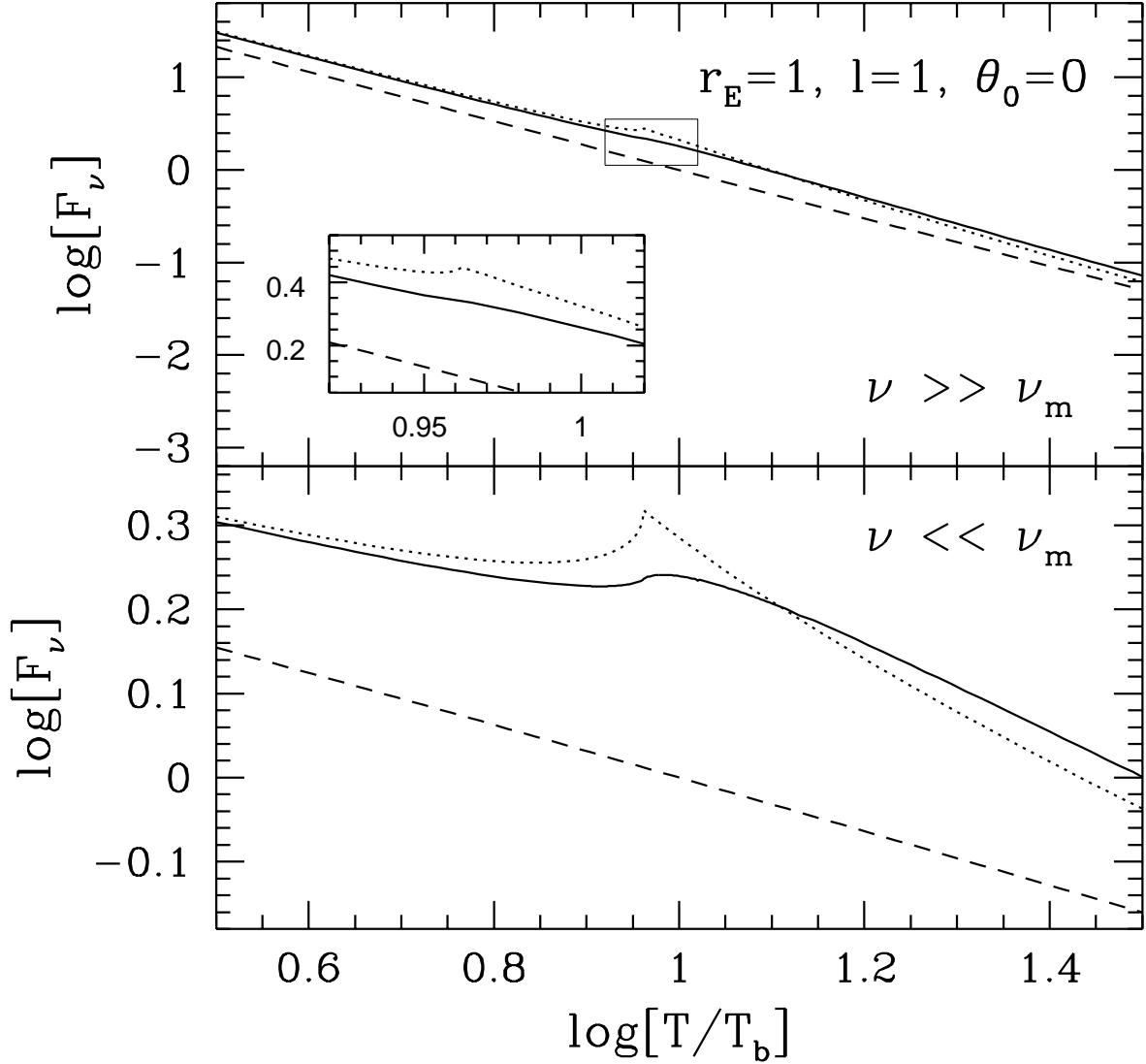}
\caption[fig3.ps]
{The lensed flux of a GRB afterglow, the unlensed flux,
and the lensed flux while retaining the initial surface brightness
at all the time as a function of the time $T$
are plotted by solid, dashed and dotted lines,
respectively.
The origin of the vertical axis has no meaning.
We set $r_E=1$, $l=1$, $\theta_0=0$ and $c_s=c/\sqrt{3}$.
The upper panel shows the flux at frequencies above the typical
synchrotron frequency,
and the lower panel is at frequencies below the typical
synchrotron frequency.
In the upper panel, the boxed region is expanded.
}
\end{figure}

\newpage 
\begin{figure}
    \epsfysize 17cm 
    \epsfbox{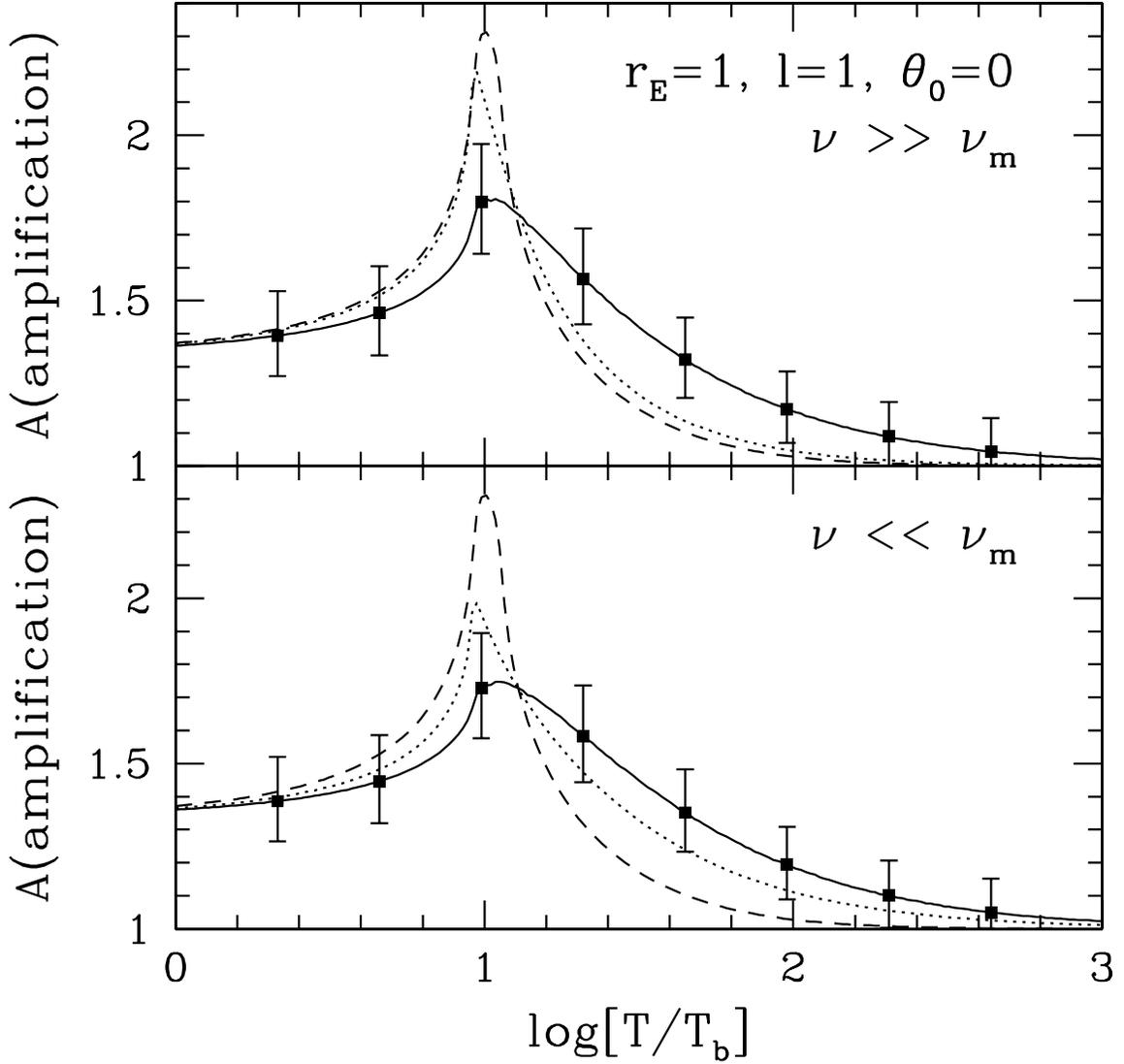}
\caption[fig4.ps]
{The amplification factor (while retaining the initial surface brightness)
as a function of the time $T$
is plotted by solid (dotted) lines.
The amplification factor for the uniform ring with a fractional width 10\%
is also plotted by dashed lines.
Errorbars of $\pm 0.1$ magnitude are also shown for reference.
We set $r_E=1$, $l=1$, $\theta_0=0$ and $c_s=c/\sqrt{3}$.
The upper panel shows the flux at frequencies above the typical
synchrotron frequency,
and the lower panel is at frequencies below the typical
synchrotron frequency.
}
\end{figure}

\newpage 
\begin{figure}
    \epsfysize 17cm 
    \epsfbox{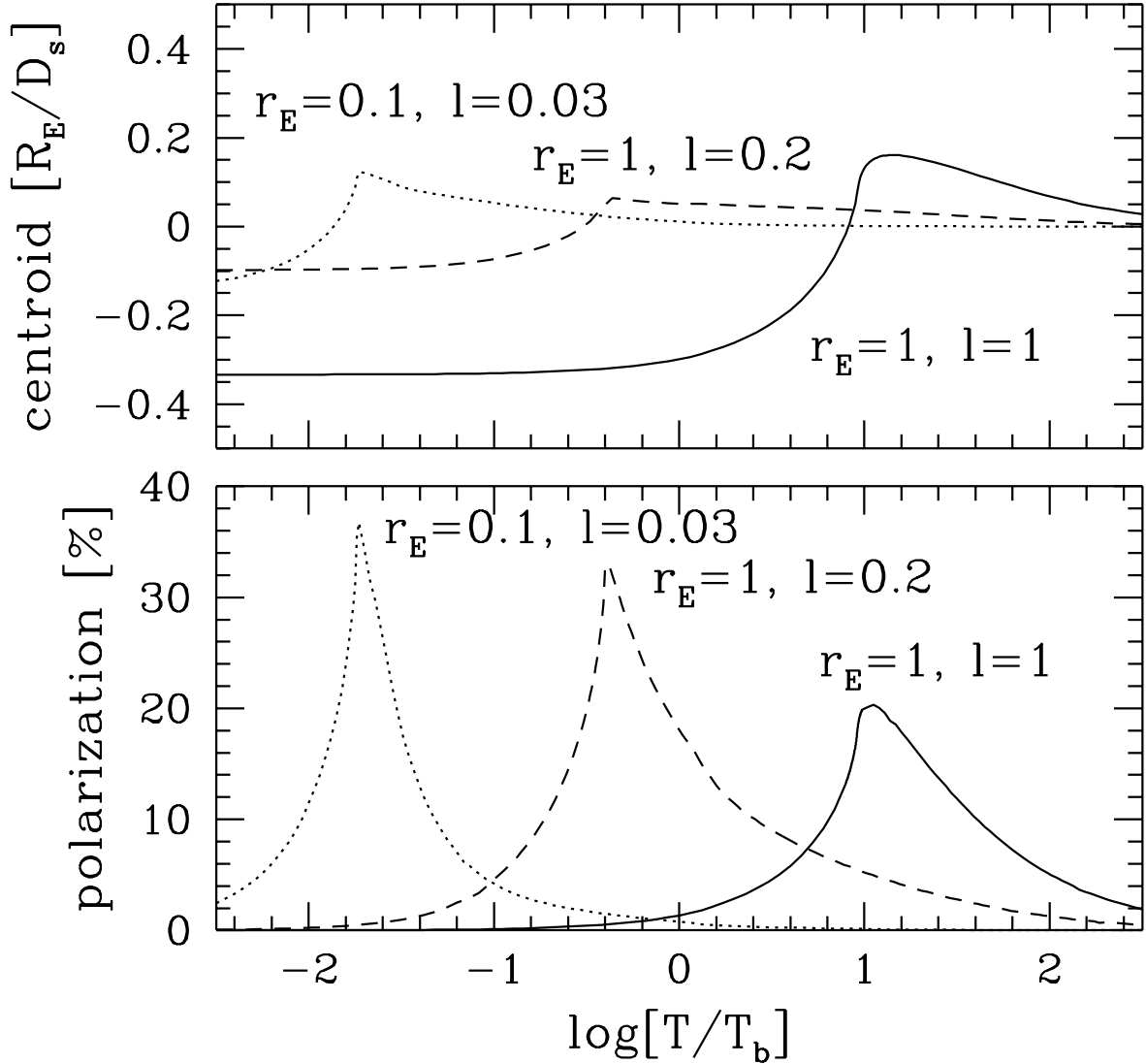}
\caption[fig5.ps]
{In the upper panel,
the light centroid position of a GRB afterglow on the sky 
in units of the ratio of the Einstein radius $R_E$ to the source distance
$D_s$ as a function of the time $T$ is shown for three cases
$(r_E, l)=(0.1, 0.03), (1, 0.2)$ and $(1, 1)$.
The horizontal axis has the same meaning as in the lower panel.
In the lower panel, the polarization of a GRB afterglow 
as a function of the observed time $T$ is shown 
for $(r_E, l)=(0.1, 0.03)$, $(1, 0.2)$ and $(1, 1)$.
The polarization scales linearly with the polarization at each point 
of the image $\Pi_0$, and we assume here $\Pi_0=(p+1)/(p+7/3)\simeq 72 \%$.
In both panels, frequencies are above the typical
synchrotron frequency.
}
\end{figure}

\newpage 
\begin{figure}
    \epsfysize 17cm 
    \epsfbox{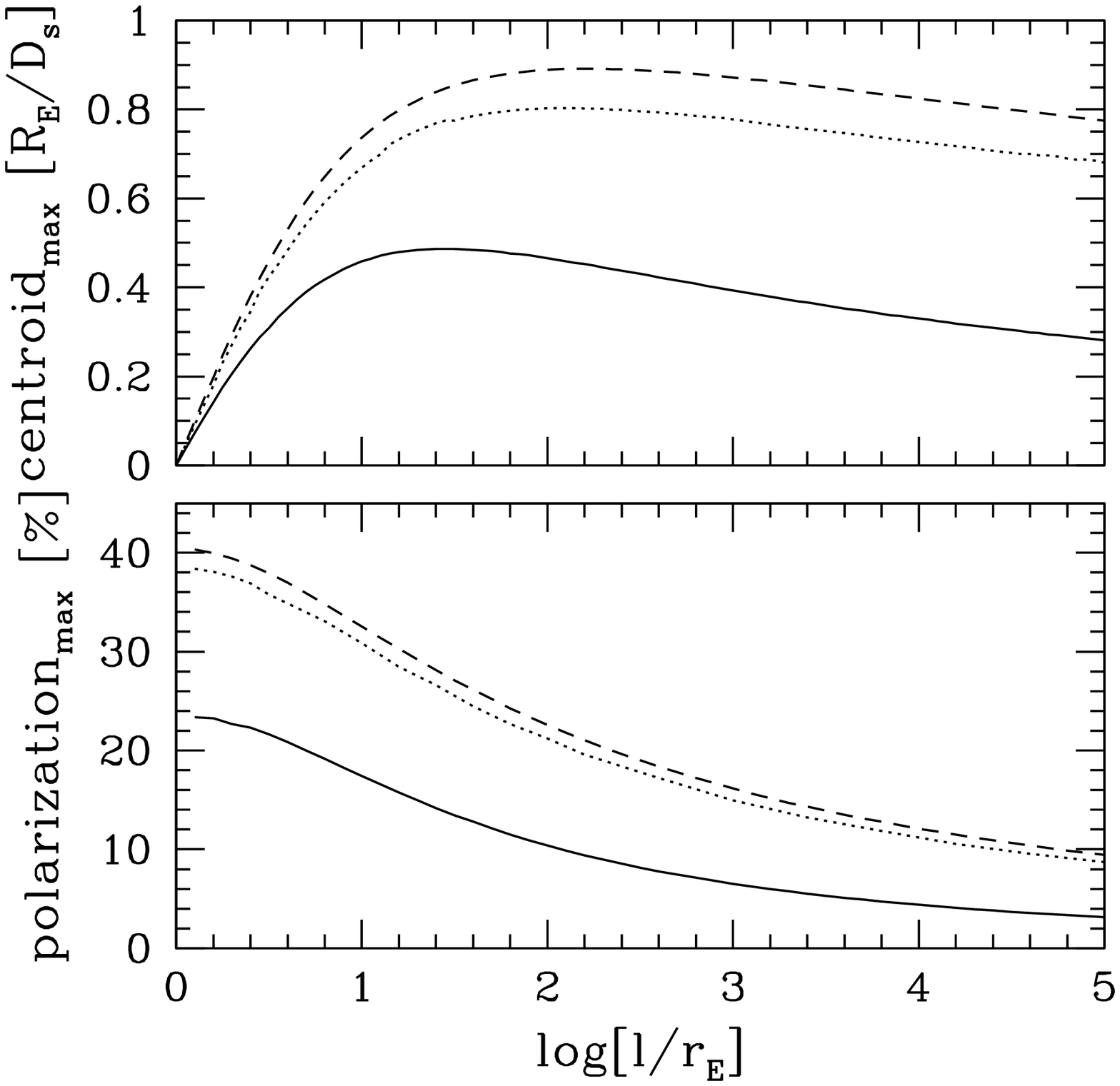}
\caption[fig6.ps]
{In the upper panel,
the maximum magnitude of the light centroid 
displacement as a function of
the normalized impact parameter $l/r_E$ is shown
for the uniform surface brightness (thick lines),
the initial surface brightness (dotted lines)
and the uniform ring with a fractional width 10\% (dashed lines).
The horizontal axis has the same meaning as in the lower panel.
In the lower panel,
the maximum polarization as a function of
the normalized impact parameter $l/r_E$
is shown for the uniform surface brightness (thick lines),
the initial surface brightness (dotted lines)
and the uniform ring with a fractional width 10\% (dashed lines).
We assume here $\Pi_0=(p+1)/(p+7/3)\simeq 72 \%$.
In both panel, the surface brightness is retained,
so that the displacement and the polarization depend only on $l/r_E$.
}
\end{figure}

%
%


\begin{references}
Alcock, C. et al. 2000, ApJ, 542, 281

Blandford, R. D., \& Mckee, C. F. 1976, Phys. Fluids, 19, 1130

Frail, D. A., et al. 2000, ApJ, 534, 559

Fukugita, M., Hogan, C. J., \& Peebles, P. J. E. 1998, ApJ, 503, 518

Garnavich, P. M., Loeb, A., \& Stanek, K. Z. 2000, ApJ, 544, L11

Gaudi, B. S., \& Loeb, A. 2001, astro-ph/0102003

Gaudi, B. S., Granot, J., \& Loeb, A. 2001, astro-ph/0105240

Ghisellini, G., \& Lazzati, D. 1999, MNRAS, 309, L7

Goodman, J. 1997, New Astron., 2, 449

Gould, A. 1995, ApJ, 455, 37

Granot, J., \& Loeb, A. 2001, ApJ, 551, L63

Granot, J., Piran, T., \& Sari, R. 1999, ApJ, 513, 679

Halpern, J. P., Kemp, J., Piran, T., \& Bershady, M. A. 1999, ApJ, 517, L105

Halpern, J. P., et al. 2000, ApJ, 543, 697

Harrison, F. A., et al. 1999, ApJ, 523, L121

H\o g, E., Novikov, I. D., \& Polnarev, A. G. 1995, A\& A, 294, 287

Hosokawa, M., Ohnishi, K., Fukushima, T., \& Takeuti, M. 1993, 
A\& A, 278, L27

Huang, Y. F., Dai, Z. G., \& Lu, T. 2000, A\&A, 355, L43

Koopmans, L. V. E., \& Wambsganss, J. 2000, astro-ph/0011029

Loeb, A., \& Perna, R. 1998, ApJ, 495, 597

Mao, S., \& Loeb, A. 2001, ApJ, 547, L97

Mao, S., \& Witt, H. J. 1998, MNRAS, 300, 1041

Miyamoto, M., \& Yoshii, Y. 1995, ApJ, 110, 1427

Moderski, R., Sikora, M., \& Bulik, T. 2000, ApJ, 529, 151

Nakamura, T. 1999, ApJ, 522, L101

Paczy${\acute {\rm n}}$ski, B. 1998, ApJ, 494, L23

Panaitescu, A. 2001, astro-ph/0102401

Panaitescu, A., \& M${\acute {\rm e}}$sz${\acute {\rm a}}$ros, P.
1998, ApJ, 493, L31

Panaitescu, A., \& M${\acute {\rm e}}$sz${\acute {\rm a}}$ros, P.
1999, ApJ, 526, 707

Press, W. H., \& Gunn, J. E. 1973, ApJ, 185, 397

Rhoads, J. E. 1999, ApJ, 525, 737

Sari, R. 1998, ApJ, 494, L49

Sari, R. 1999, ApJ, 524, L43

Sari, R., Piran, T., \& Halpern, J. P. 1999, ApJ, 519, L17

Sari, R., Piran, T., \& Narayan, R. 1998, ApJ, 497, L17

Schneider, P., Ehlers, J., \& Falco, E. E. 1992,
Gravitational Lenses (Berlin: Springer)

Stanek, K. Z., et al. 1999, ApJ, 522, L39

Walker, M. A. 1995, ApJ, 453, 37

Waxman, E. 1997, ApJ, 491, L19

\end{references}
\end{document}